\documentclass[a4paper,12pt]{article}
\def\be {\begin{equation}}
\def\ee {\end{equation}}
\def\ba {\begin{eqnarray}}
\def\ea {\end{eqnarray}}

\def\bea{\begin{eqnarray}}
\def\eea{\end{eqnarray}}
\def\bi {\begin{itemize}}
\def\ei {\end{itemize}}
\begin{document}

\title{\bf \large {Thermodynamical Interpretation of the Interacting
Holographic Dark Energy Model in a non-flat Universe}}
\author{\normalsize{M. R. Setare$^{1}$\thanks{%
E-mail: rezakord@ipm.ir}  \, and\, Elias C. Vagenas$^{2}$\thanks{%
E-mail: evagenas@academyofathens.gr} }\\
\newline
\\
{\normalsize \it $^1$ Department of Science, Payame Noor
University, Bijar, Iran}
\\
{\normalsize \it $^2$
 Research Center for Astronomy \& Applied Mathematics,}\\
 {\normalsize \it Academy of Athens,
 Soranou Efessiou 4,
 GR-11527, Athens, Greece}}

\date{\small{}}

\maketitle
\begin{abstract}
Motivated by the recent work of Wang, Lin, Pavon, and Abdalla
\cite{1}, we generalize their work to the non-flat case. In
particular, we provide a thermodynamical interpretation for the
holographic dark energy model in a non-flat universe. For this case,
the characteristic length is no more the radius of the event horizon
($R_E$) but the event horizon radius as measured from the sphere of
the horizon ($L$). Furthermore, when interaction between the dark
components of the holographic dark energy model in the non-flat
universe is present its thermodynamical interpretation changes by a
stable thermal fluctuation. A relation between the interaction term
of the dark components and this thermal fluctuation is obtained. In
the limiting case of a flat universe, i.e. $k=0$, all results given
in \cite{1} are obtained.
 \end{abstract}

\newpage

\section{Introduction}
Recent observations from type Ia supernovae \cite{SN} associated
with Large Scale Structure \cite{LSS} and Cosmic Microwave
Background anisotropies \cite{CMB} have provided main evidence for
the cosmic acceleration. The combined analysis of cosmological
observations suggests that the universe consists of about $70\%$
dark energy, $30\%$ dust matter (cold dark matter plus baryons),
and negligible radiation. Although the nature and origin of dark
energy are unknown, we still can propose some candidates to
describe it, namely  since we do not know where this dark energy comes from,
and how to compute it from the first principles, we search for
phenomenological models. The astronomical observations will then select
one of these models.
The most obvious theoretical candidate of dark energy is the
cosmological constant $\lambda$ (or vacuum energy)
\cite{Einstein:1917,cc} which has the equation of state parameter $w=-1$.
However, as it is well known, there are two difficulties that arise
from the cosmological constant scenario, namely the two famous
cosmological constant problems --- the ``fine-tuning'' problem and
the ``cosmic coincidence'' problem \cite{coincidence}. An
alternative proposal for dark energy is the dynamical dark energy
scenario. This  dynamical proposal is often realized by
some scalar field mechanism which suggests that the specific energy form
with negative pressure is provided by a scalar field evolving down
a proper potential. So far, a plethora of scalar-field dark
energy models have been studied, including quintessence
\cite{quintessence}, K-essence \cite{kessence}, tachyon
\cite{tachyon}, phantom \cite{phantom}, ghost condensate
\cite{ghost2} and quintom \cite{quintom}, and so forth.
It should be noted that the mainstream viewpoint regards the scalar-field
dark energy models as an effective description of an underlying
theory of dark energy. In addition, other proposals on dark energy
include interacting dark energy models \cite{intde}, braneworld
models \cite{brane}, Chaplygin gas models \cite{cg},
and many others.\\
Currently, an interesting attempt for probing the nature of dark
energy within the framework of quantum gravity
(and thus compute it from first principles) is the so-called ``Holographic
Dark Energy'' (HDE) proposal \cite{Cohen:1998zx,Horava:2000tb,Hsu:2004ri,Li:2004rb}.
It is well known that the holographic principle is an important result of the
recent researches for exploring the quantum gravity (or string
theory) \cite{holoprin}.  The HDE model has been
tested and constrained by various astronomical observations
\cite{Huang:2004wt,obs3} as well as by the Anthropic Principle
\cite{Huang:2004mx}. Furthermore, the HDE model
has been extended to include the spatial curvature contribution,
i.e. the HDE model in non-flat space
\cite{nonflat}. For other
extensive studies, see e.g. \cite{holoext}.\\
It is known that the coincidence or, ``why now" problem is easily
solved in some models of HDE based on the
fundamental assumption that matter and holographic dark energy do
not conserve separately \cite{interac,Amendola:2000uh}. In
fact a suitable evolution of the Universe is obtained when, in
addition to the holographic dark energy, an interaction (decay of
dark energy to matter) is assumed.\\
%
%
%
Since we know neither the nature of dark energy nor the nature of
dark matter, a microphysical interaction model is not available
either. However, pressureless dark matter in interaction with
holographic dark energy is more than just another model to
describe an accelerated expansion of the universe. It provides a
unifying view of  different models which are viewed as different
realizations of the Interacting HDE Model at
the perturbative level \cite{Zimdahl:2007ne}.
Since the discovery of black hole thermodynamics in 1970, physicists
have speculated on the thermodynamics of cosmological models in an
accelerated expanding universe \cite{thermo}. Related to the present
work, for time-independent and time-dependent
equations of state (EoS), the first and second laws of
thermodynamics in a flat universe were investigated in \cite{abdalla}.
In particular, for the case of a constant EoS,
the first law of thermodynamics is valid for the apparent horizon
(Hubble horizon) but it does not hold for the event horizon when viewed as
system's IR cut-off. When the EoS is assumed to be time-dependent,
using a holographic model of dark energy in flat space, the same
result is obtained: the event horizon, in contrast to the apparent
horizon, does not satisfy the first law. Additionally, while the event
horizon does not respect the second law of thermodynamics,
it holds for the universe enclosed by the apparent horizon.
\par\noindent
In the present paper we extend the work by Wang, Lin, Pavon, and
Abdalla \cite{1} to the interacting HDE model of dark energy in a
non-flat universe, we study the thermodynamical interpretation of
the interacting holographic dark energy model for a universe
enveloped by the event horizon measured from the  sphere of the
horizon named $L$. The remainder of the paper is as follows. In
Section 2 we generalize the thermodynamical picture of the
non-interacting HDE model in a non-flat universe. In Section 3, we
extend the thermodynamical picture in the case where there is an
interaction term between the dark components of the HDE model. An
expression for the interaction term in terms of a thermal
fluctuation is given. In the limiting case of flat universe,we
obtain the results derived  in \cite{1}. Finally, Section 4 is
devoted to concluding remarks.
\section{Thermodynamical Picture of the non-Interacting HDE model}
\par\noindent
In this section we consider the HDE model when
there is no interaction between the holographic energy density
$\rho_{X}$ and the Cold Dark Matter (CDM) $\rho_{m}$ with $w_{m}=0$.
In addition, non-dark components have been considered negligible and thus are not included.
The third Friedmann equation describes the time evolution of the energy densities
of the dark components. These equations are actually the continuity equations for the dark energy and CDM
\begin{eqnarray}
\label{2eq1}&& \dot{\rho}_{X}+3H(1+w_{X}^{0})\rho_{X} =0, \\
\label{2eq2}&& \dot{\rho}_{m}+3H\rho_{m} =0
\end{eqnarray}
where the quantity $H=\dot{a}/a$ is the Hubble parameter and the superscript above
the equation of state parameter, $w_{X}$, denotes that there is no interaction between the dark components.
The non-interacting HDE model will be accommodated in the non-flat Friedmann-Robertson-Walker universe
which is described by the line element
 \be\label{metr}
ds^{2}=-dt^{2}+a^{2}(t)(\frac{dr^2}{1-kr^2}+r^2d\Omega^{2})
 \ee
where $a=a(t)$ is the scale factor of the non-flat Friedmann-Robertson-Walker universe
and $k$ denotes the curvature of space with $k=0,\,1,\,-1$ for flat,
closed and open universe, respectively. A closed universe with a
small positive curvature ($\Omega_{k}\sim 0.01$) is compatible with
observations \cite{ {wmap}, {ws}}. Thus, in order to connect
the curvature of the universe to the energy density, we employ the
first Friedmann equation given by
\begin{equation}
\label{2eq7} H^2+\frac{k}{a^2}=\frac{1}{3M^2_p}\Big[
 \rho_{X}+\rho_{m}\Big]
\end{equation}
where $c$ is a positive constant in the HDE model and $M_p$
is the reduced Planck mass.
We also define the dimensionless density parameters
\begin{equation}
\label{2eq9} \Omega_{m}=\frac{\rho_{m}}{\rho_{cr}}=\frac{
\rho_{m}}{3M_p^2H^2} \hspace{1ex},\hspace{1cm}
\Omega_{X}=\frac{\rho_{X}}{\rho_{cr}}=\frac{ \rho_{X}}{3M^2_pH^2}
\hspace{1ex},\hspace{1cm} \Omega_{k}=\frac{k}{a^2H^2} \hspace{1ex}.
\end{equation}
\par\noindent
Therefore, we can rewrite the first Friedmann equation as
\begin{equation} \label{2eq10} \Omega_{m}+\Omega_{X}-\Omega_{k}=1\hspace{1ex}.
\end{equation}
For completeness, we give the deceleration parameter%
\be
\label{qequ}
q=-\frac{\ddot{a}}{H^2a}=-\left(\frac{\dot{H}}{H^2}+1\right)
\hspace{1ex}
\ee
which combined with the Hubble parameter and the
dimensionless density parameters form a set of useful parameters for
the description of the astrophysical observations.
It should be stressed that in the non-flat universe
the characteristic length which plays the role of the IR-cutoff is
the radius $L$ of the event horizon measured on the sphere of the
horizon and not the radius $R_{h}$ measured on the radial direction.
Therefore, the holographic dark energy density is given as
\be
\label{density1}
\rho_{X}=\frac{3c^{2}M^{2}_{p}}{L^{2}}
\hspace{1ex}.
\ee
The radius $L$ is given by
\be
\label{radius1}
L=a r(t)
\ee
where the function $r(t)$ is defined through the equation
\be
\label{radius2}
\int_{0}^{r(t)}\frac{dr}{\sqrt{1-k
r^2}}=\frac{R_{h}}{a}
\hspace{1ex}.
\ee
Solving for the general case
of non-flat universe the above equation, the function $r(t)$ is
given as
\be
\label{radius3}
r(t)=\frac{1}{\sqrt{k}}\sin y
\ee
where
\be
\label{argument}
y=\frac{\sqrt{k} R_{h}}{a}
\hspace{1ex}.
\ee
Substituting equation (\ref{density1}) in the expression for the
dimensionless density parameter of the holographic dark energy as
given by equation (\ref{2eq9}), one gets
\be
\label{HL1}
HL=\frac{c}{\sqrt{\Omega_{X}^{0}}}
\ee
and thus
\be
\label{HL2}
\dot{L}=HL+a\dot{r}(t)=\frac{c}{\sqrt{\Omega_{X}^{0}}}-\cos y
\hspace{1ex}.
\ee
Differentiating the holographic dark energy
density as given by equation (\ref{density1}) and using equations
(\ref{HL1}) and (\ref{HL2}), one gets
\be
\label{density2}
\dot{\rho}_{X}=-2H\left(1-\frac{\sqrt{\Omega_{X}^{0}}}{c}\cos
y\right)\rho_{X}
\ee
and thus the conservation equation for the
holographic dark energy (\ref{2eq1}) yields
\be
\label{eosp1}
1+3\omega_{X}^{0}=-2\frac{\sqrt{\Omega_{X}^{0}}}{c}\cos y
\hspace{1ex}.
\ee
Following \cite{1} (see also \cite{Pavon:2007gt}), the non-interacting
HDE model in the non-flat universe as described
above is thermodynamically interpreted as a state in thermodynamical
equilibrium. According to the generalization of the black hole
thermodynamics to the thermodynamics of cosmological models, we have taken the
temperature of the event horizon to be $T_L=(1/2\pi L)$ which is actually the only
temperature to handle in the system.
If the fluid temperature of the cosmological model is set
equal to the horizon temperature ($T_L$), then the system will be
in equilibrium. Another possibility \cite{davies2} is that the fluid
temperature is proportional to the horizon temperature, i.e. for the
fluid enveloped by the apparent horizon $T=eH/2\pi$ \cite{pavon}.
In general, the systems must interact for some length of time before
they can attain thermal equilibrium. In the case at hand, the
interaction certainly exists as any variation in the energy density
and/or pressure of the fluid will automatically induce a
modification of the horizon radius via Einstein's equations.
Moreover, if $T \neq T_{L}$, then energy would spontaneously flow
between the horizon and the fluid (or viceversa), something at
variance with the FRW geometry \cite{pa}. Thus, when we
consider the thermal equilibrium state of the universe, the
temperature of the universe is associated with the horizon temperature. In this
picture the equilibrium entropy of the holographic dark energy is
connected with its energy and pressure through the first
thermodynamical law
\be
\label{law1} TdS_{X}=dE_{X}+p_{X}dV
\ee
where the volume is given as
\be
V=\frac{4\pi}{3}L^{3} \hspace{1ex},
\ee
the energy of the holographic dark energy is defined as
\be
\label{energy1}
E_{X}=\rho_{X} V=4\pi c^{2} M^{2}_{p}L
\ee
and the temperature of the event horizon is given as
\be
\label{temp1}
T=\frac{1}{2\pi L^{0}}
\hspace{1ex}.
\ee
Substituting the aforesaid expressions for the volume, energy, and temperature in equation
(\ref{law1}) for the case of the non-interacting HDE model, one obtains
\be
\label{entropy1}
dS_{X}^{(0)}=8\pi^{2}c^{2}M^{2}_{p}\left(1+3\omega^{0}_{X}\right)L^{0}dL^{0}
\ee
and implementing equation (\ref{eosp1}) the above-mentioned
equation takes the form
\be
\label{entropy2}
dS_{X}^{(0)}=-16\pi^{2}c M^{2}_{p}\sqrt{\Omega^{0}_{X}}\cos y\,
L^{0}dL^{0}
\ee
where the superscript $(0)$ denotes that in this
thermodynamical picture our universe is in a thermodynamical stable
equilibrium.
\par\noindent
In the case of flat universe, i.e. $k=0$, we obtain
\be
dS_{X}^{(0)}=-16\pi^{2}c M^{2}_{p}\sqrt{\Omega^{0}_{X}}\,L^{0}dL^{0}
\ee
which is exactly the result derived in \cite{1} when one replaces $L^{0}$ with the
future event horizon $R^{0}_{E}$.
\section{Thermodynamical Picture of the Interacting HDE model}
\par\noindent
In this section we consider the HDE model when
there is interaction between the holographic energy density $\rho_{X}$ and the Cold Dark Matter (CDM) $\rho_{m}$.
The corresponding continuity equations are now written as
\begin{eqnarray}
\label{eq3}&&
\dot{\rho}_{X}+3H(1+w_{X})\rho_{X} =-Q, \\
\label{eq4}&&
\dot{\rho}_{m}+3H\rho_{m} = Q
\end{eqnarray}
where the quantity $Q$ expresses the interaction between the dark components. The
interaction term $Q$ should be positive, i.e. $Q>0$, which means that there is an energy transfer from
the dark energy to dark matter. The positivity of the interaction term ensures that the second law of
thermodynamics is fulfilled \cite{Pavon:2007gt}.
At this point, it should be stressed that the continuity equations imply that the
interaction term should be a function of a quantity with units of
inverse of time (a first and natural choice can be the Hubble factor $H$)
multiplied  with the energy density. Therefore, the interaction term could be
in any of the following forms: (i) $Q\propto H\rho_{X}$ \cite{Pavon:2005yx,Pavon:2007gt},
(ii) $Q\propto H\rho_{m}$ \cite{Amendola:2006dg}, or
(iii) $Q\propto H(\rho_{X}+\rho_{m})$ \cite{Wang:2005ph}. The freedom of
choosing the specific form of the interaction term $Q$  stems from our
incognizance of the origin and nature of dark energy as well as dark matter.
Moreover, a microphysical model describing the interaction between the dark components
of the universe is not available nowadays.

The interacting HDE model will again be accommodated in the non-flat
Friedmann-Robertson-Walker universe described by the line element
(\ref{metr}). Our analysis here will give same results with those in
the non-interacting case concerning the first Friedmann equation,
dimensionless density parameters, and the characteristic length as
well as equations related to them (see  equations
(\ref{2eq7})\hspace{1ex}-\hspace{1ex}(\ref{density2})). However, due
to the existence of interaction between the dark components of the
holographic dark energy model which changed the conservation
equations, equation (\ref{eosp1}) derived for the non-interacting
HDE model has to be changed accordingly. Thus,
by substituting equation (\ref{density2}) in the conservation
equation (\ref{eq3}) for the dark energy component one obtains
\be
\label{eosp2}
1+3\omega_{X}=-2\frac{\sqrt{\Omega_{X}}}{c}\cos y -
\frac{Q}{3H^{3}M^{2}_{p}\Omega_{X}}
\hspace{1ex}.
\ee
Comparing equation (\ref{eosp2}) with equation (\ref{eosp1}), it is easily
seen that the presence of the interaction term $Q$ has provoked a
change in the equation of state parameter and consequently in the
dimensionless density parameter of the dark energy component and
thus now there is no subscript above the aforesaid quantities to
denote the absence of interaction. According to \cite{1}, the
interacting HDE model in the non-flat universe
as described above is not anymore thermodynamically interpreted as a
state in thermodynamical equilibrium. In this picture the effect of
interaction between the dark components of the HDE model is thermodynamically interpreted as a small fluctuation
around the thermal equilibrium. Therefore, the entropy of the
interacting holographic dark energy is connected with its energy and
pressure through the first thermodynamical law
\be
\label{law2}
TdS_{X}=dE_{X}+p_{X}dV
\ee
where now the entropy has been assigned
an extra logarithmic correction \cite{das}
\be
S_{X}=S_{X}^{(0)}+S_{X}^{(1)}
\ee
where
\be
\label{correction1}
S_{X}^{(1)}=-\frac{1}{2}\ln \left(C T^{2}\right)
\ee
and $C$ is the heat capacity defined by
\be
C=T\frac{\partial S_{X}^{(0)}}{\partial
T}
\ee
and using equations (\ref{entropy1}), (\ref{temp1}), and
(\ref{eosp1}) is given as
\bea
\label{capacity1}
C&\hspace{-1ex}=\hspace{-1ex}&-8\pi^{2}c^{2}M^{2}_{p}(L^{0})^{2}(1+3\omega_{X}^{0})\\
\label{capacity2}
&\hspace{-1ex}=\hspace{-1ex}&16\pi^{2}c M^{2}_{p}(L^{0})^{2} \sqrt{\Omega_{X}^{0}} \cos y
\hspace{1ex}.
\eea
Substituting the expressions for the volume, energy, and temperature (it is noteworthy that
these quantities depend now on $L$ and not on $L^{0}$ since there is interaction among the dark components)
in equation (\ref{law2}) for the case of the interacting HDE model,
one obtains
\be
\label{entropy3}
dS_{X}=8\pi^{2}c^{2}M^{2}_{p}\left(1+3\omega_{X}\right)L dL
\ee
and thus one gets
\bea
\label{eosp3}
1+3\omega_{X}&\hspace{-1ex}=\hspace{-1ex}&\frac{1}{8\pi^{2}c^{2}M^{2}_{p}L}\frac{dS_{X}}{dL}\\
&\hspace{-1ex}=\hspace{-1ex}&\frac{1}{8\pi^{2}c^{2}M^{2}_{p}L}\left[\frac{dS_{X}^{(0)}}{dL}+\frac{dS_{X}^{(1)}}{dL}\right]\\
\label{eosp4}
&\hspace{-1ex}=\hspace{-1ex}&-2\left(\frac{\sqrt{\Omega_{X}^{0}}}{c}\cos y \right)\frac{L^{0}}{L}
\frac{dL^{0}}{dL}+\frac{1}{8\pi^{2}c^{2}M^{2}_{p}L}\frac{dS_{X}^{(1)}}{dL}
\eea
where the last term concerning the logarithmic correction can be computed using
expressions (\ref{correction1}) and (\ref{capacity2})
\be
\frac{dS_{X}^{(1)}}{dL}=-\frac{H}{\left(\frac{c}{\sqrt{\Omega_{X}^{0}}}-\cos y\right)}
\left[\frac{\left(\Omega_{X}^{0}\right)'}{4\Omega_{X}^{0}}+ y \tan y\right]
\ee
with the prime $(\hspace{1ex}')$ to denote differentiation with respect to $\ln a$.
\par\noindent
Therefore, by equating the expressions (\ref{eosp2}) and (\ref{eosp4}) for the equation of state parameter of the holographic
dark energy evaluated on cosmological and thermodynamical grounds respectively, one gets an expression for
the interaction term
\be
\label{interaction}
\frac{Q}{9H^{3}M^{2}_{p}}=
\frac{\Omega_{X}}{3}\left[-\frac{2\sqrt{\Omega_{X}}}{c}\cos y +\left(\frac{2\sqrt{\Omega_{X}}}{c}\cos y\right)\frac{L^{0}}{L}\frac{dL^{0}}{dL}\right]-
\frac{1}{8\pi^{2}c^{2}M^{2}_{p}L}\frac{\Omega_{X}}{3}\frac{dS_{X}^{(1)}}{dL}
\hspace{1ex}.
\ee
It is noteworthy that in the limiting case of flat universe, i.e. $k=0$,
we obtain exactly the result derived in \cite{1} when one replaces $L^0$ and $L$
with $R^{0}_E$ and $R_E$, respectively.
\section{Conclusions}
\par\noindent
Understanding dark energy is one of the biggest challenges to
the particle physics of this century. Studying the interaction between
the dark energy and ordinary matter will open a possibility of
detecting the dark energy. It should be pointed out that evidence was
recently provided by the Abell Cluster A586 in support of the
interaction between dark energy and dark matter
\cite{Bertolami:2007zm}. However, despite the fact that numerous
works have been performed till now, there are no strong
observational bounds on the strength of this interaction
\cite{Feng:2007wn}. This weakness to set stringent
(observational or theoretical) constraints on the strength of the
coupling between dark energy and dark matter stems from our
unawareness of the nature and origin of dark components of the
Universe. It is therefore more than obvious that further work is
needed to this direction.\\
In 1973, Bekenstein \cite{bek} assumed that there is a relation
between the area of the event horizon of a black hole and the thermodynamics of a black hole,
so that the area of the event horizon of the black hole is a measure of the
black hole entropy. Along this line of thought, it was argued in
\cite{jac} that the gravitational Einstein equations can be derived
through a thermodynamical argument using the relation between area
and entropy as input. Following \cite{{jac},{fro}}, Danielsson
\cite{dan} has been able to obtain the Friedmann equations, by
applying the relation $\delta Q=T dS$ to a cosmological horizon and
calculate the heat flow through the horizon of an expanding universe
in an acceleration phase.
 This idea has been generalized to horizons of
cosmological models, so that each horizon corresponds to an entropy.
Therefore, the second law of thermodynamics was modified in a way that
in its generalized form, the sum of all time derivatives of entropies
related to horizons plus the time derivative of normal entropy must be
positive i.e. the sum of entropies must be an increasing function of
time.\\
 In the present paper, we have provided a thermodynamical interpretation for the
HDE model in a non-flat universe. We utilized the horizon's radius $L$
measured from the sphere of the horizon as the system's IR cut-off.
We investigated the thermodynamical picture of the interacting HDE model for a non-flat
universe enveloped by this horizon. The non-interacting HDE model in a non-flat universe was
thermodynamically interpreted as a thermal equilibrium state.
When an interaction between the dark components of the HDE model in the non-flat
universe was introduced the thermodynamical interpretation of the HDE model changed.
The thermal equilibrium state was perturbed by a stable thermal fluctuation which was now the
 thermodynamical interpretation
of the interaction. Finally, we have derived an expression that
connects this interaction term of the dark components of the interacting HDE model in a non-flat universe
with the aforesaid thermal fluctuation.

\end{document}